\documentclass[twocolumn,showpacs,prl]{revtex4}
\usepackage{graphicx}
\begin{document}

\title{Formation of a molecular Bose-Einstein condensate and an
entangled atomic gas by Feshbach resonance}

\author{V. A. Yurovsky}

\author{A. Ben-Reuven}

\affiliation{School of Chemistry, Tel Aviv University, 69978 Tel Aviv,
Israel}

\date{\today}
\begin{abstract}

Processes of association in an atomic Bose-Einstein condensate,
and dissociation of the resulting molecular condensate, due to
Feshbach resonance in a time-dependent magnetic field, are analyzed
incorporating non-mean-field quantum corrections and inelastic
collisions. Calculations for the Na atomic condensate demonstrate
that there exist optimal conditions under which about 80\% of the
atomic population can be converted to a relatively long-lived
molecular condensate (with lifetimes of 10 ms and more). Entangled
atoms in two-mode squeezed states (with noise reduction of about 30
dB) may also be formed by molecular dissociation. A gas of atoms in
squeezed or entangled states can have applications in quantum
computing, communications, and measurements.

\end{abstract}
\maketitle    

\paragraph*{Introduction.---}

The recently discovered Bose-Einstein condensates (BEC), or
matter waves, resemble in certain ways coherent electromagnetic
radiation. This similarity stimulated the development of atomic optics
\cite{M01}, involving non-classical states of the atomic fields, such
as squeezed and entangled states \cite{SZ97}. Squeezed states are
characterized by noise reduction, and can be applied in communications
and measurements. Entangled states of a decomposable system cannot be
expressed as a product of the component states, and can be used in
quantum computing and communications. Squeezed atomic states can be
formed in four-wave mixing \cite{D99}, in arrays of atomic traps
\cite{OTFYK01}, in multimode condensates \cite{SDCZ01,DBE01}, in the
decay of unstable BEC \cite{Y02}, in collisions of BEC wavepackets
\cite{Y02}, and as the outcome of Bogolubov fluctuations subject to
stimulated light scattering \cite{RGB01} or Belyaev dumping
\cite{RCNB01}. The squeezing can be measured experimentally by using
homodyne detection, analogous to the one used in quantum optics (see
Ref.\ \cite{SZ97}). The key component of this method --- a beam
splitter --- already exists (see Ref.\ \cite{BeamSpl}).

The present work suggests the dissociation of molecular BEC as a
source of atom fields in two-mode squeezed states that are entangled.
Formation of single-mode squeezed states by the same mechanism has
been discussed in  Refs.\ \cite{PM01,VYA01,YBJ02}. Formation of
entangled atomic pairs in the dissociation of individual diatomic
molecules has been considered in Ref.\ \cite{OK01}. Other mechanisms
of formation of entangled gases have been discussed in Refs.\
\cite{SDCZ01,Y02,RGB01}.

The molecular BEC required as the source of the entangled gas is
interesting in its own right, although it has not been realized yet.
Formation of a molecular BEC by direct cooling of molecular gases is
obstructed by the rotational degrees of freedom. An alternative method
is the association of atomic BEC \cite{Photoassoc,TTHK99}. A process
of photoassociation \cite{Photoassoc}, realized experimentally
\cite{WFHRH00}, is obstructed by spontaneous emission \cite{JM02}. We
consider here the association of atoms in BEC by Feshbach resonance
\cite{TTHK99} in a time-dependent magnetic field. Such a process is
associated with a large condensate loss observed in recent experiments
\cite{Feshbach_exp}. This loss follows from deactivation of the
resonant molecules by inelastic collisions \cite{YBJW99,TTHK99}, as
well as from the formation of non-condensate atoms by molecular
dissociation. An advantage of the use of Feshbach association is the
possibility of reducing the negative effect of collisions by lowering
the condensate density. Although the molecular BEC is made of excited
molecules, and is therefore unstable, it still can be used as a source
of atoms in entangled and squeezed states produced by the
dissociation.

Reference\ \cite{MTJ00} treats the condensate loss as a
dissociation of single molecules. Many-body effects have been
incorporated in Ref.\ \cite{YBJW99} by introducing a width to the
molecular condensate state. A more rigorous analysis has been
performed in Ref.\ \cite{HPW01}, using second-order correlation
functions. However, it was limited to the case of a time-independent
zero detuning between the atomic and molecular states, and
deactivating collisions were not taken into account. As in the present
analysis, Ref.\ \cite{YBJW99} dealt with a time-dependent crossing of
these states, in accordance with the experiments \cite{Feshbach_exp},
and took into account deactivating collisions. We generalize here the
parametric approximation, used in Refs.\ \cite{VYA01,YBJ02} for the
description of molecular dissociation into a single atomic mode, and
in Ref.\ \cite{Y02} for a multimode analysis of fluctuations in an
unstable BEC.

The most impressive outcomes of the present approach are the
extent of (near-total) conversion to entangled atoms or a molecular
condensate, as well as the extreme degree of squeezing and the
relatively long molecular BEC lifetimes achievable.

\paragraph*{The model.---}

Consider a system of coupled atomic and molecular fields (see
Ref.\ \cite{YBJW99}) described by annihilation operators in the
momentum representation $\hat{\Psi }_{a}\left( {\bf p},t\right) $ and
 $\hat{\Psi }_{m}\left( {\bf p},t\right) $, respectively. The
coupling of the atomic and molecular fields (see Refs.\
\cite{YBJW99,TTHK99}) contains a product of two atomic creation
operators and therefore describes the formation of entangled atomic
pairs, in analogy with parametric down-conversion in quantum optics
(see Ref.\ \cite{SZ97}). Spatial inhomogeneity due to the trapping
potential and elastic collisions can be neglected here (see discussion
below). Unlike the mean field used in Ref.\ \cite{YBJW99}, the atomic
field is treated here as second-quantized, as in Ref.\ \cite{YBJ02}.

Let the initial state of the atomic field at $t=t_{0}$  be a coherent
state of zero kinetic energy
\begin{equation}
\hat{\Psi }_{a}\left( {\bf p},t_{0}\right) |\text{in}\rangle =\left(
 2\pi \right) ^{3/2}\varphi _{0}\delta \left( {\bf p}\right)
|\text{in}\rangle  ,
\end{equation}
where $|\varphi _{0}|^{2}=n_{a}\left( t_{0}\right) $ is the initial
 atomic density and $|$in$\rangle $ is the
time-independent state vector in the Heisenberg representation. A pair
of condensate atoms forms a molecule of zero kinetic energy. Therefore
the resonant molecules can be represented by a mean field $\varphi
 _{m}\left( t\right) $, such
that
\begin{equation}
\langle \text{in}|\hat{\Psi }_{m}\left( {\bf p},t\right)
|\text{in}\rangle =\left( 2\pi \right) ^{3/2}\varphi _{m}\left(
 t\right) \delta \left( {\bf p}\right)  ,
\end{equation}
where $|\varphi _{m}\left( t\right) |^{2}=n_{m}\left( t\right) $ is
 the molecular condensate density. This
approach (unlike Refs.\ \cite{Y02,YBJ02}) takes into account the time
dependence of the molecular mean field. Fluctuations of the molecular
field due to collisions involving non-condensate atoms are neglected.

The outcome of atom-molecule and molecule-molecule deactivating
collisions is introduced, as in Ref.\ \cite{YBJW99}, by adding
molecular ``dump'' states. The elimination of these states in a
second-quantized description is, however, different. It is similar to
the Heisenberg-Langevin formalism in quantum optics (see Ref.\
\cite{SZ97}), but takes into account the nonlinearity of the
collisional dumping. In the Markovian approximation, the equation of
motion for the atomic field attains the form
\begin{equation}
i\dot{\hat{\Psi }}_{a}\left( {\bf p},t\right)  =H\hat{\Psi }_{a}\left
( {\bf p},t\right)  +2g^{*}\varphi _{m}\left( t\right) \hat{\Psi
 }^{\dag }_{a}\left( -{\bf p},t\right)  +i \hat{F}\left( {\bf
 p},t\right)  \label{Psia}
\end{equation}
(using units with $\hbar =1$), where
\begin{equation}
H={p{ } ^{2}\over 2m} -\mu  {B\left( t\right) -B{ } _{0}\over 2}
-i\gamma |\varphi _{m}\left( t\right) |^{2}
\end{equation}
and $m$ is the atomic mass. The second term in $H$ describes the
time-dependent Zeeman shift of the atom in an external magnetic field
$B\left( t\right) $, relative to half the energy of the molecular
 state, which is
chosen as the zero energy, $\mu $ is the difference in magnetic
 momenta of
an atomic pair and a molecule, and $B_{0}$  is the resonance value of
 $B$.
The coupling of the atomic and the molecular fields $g$ is related to
the phenomenological resonance strength $\Delta $ as $|g|^{2}=2\pi
|a_{a}\mu |\Delta /m$ (see
Ref.\ \cite{YBJW99}), where $a_{a}$  is the elastic scattering
 length. The
parameter $\gamma $  describes the width of atomic states due to
 deactivating
collisions (see Ref.\ \cite{YBJW99}). The quantum noise source
 $\hat{F}\left( {\bf p},t\right) $
provides conserving the correct commutation relations of the field
operators.

Generalizing the parametric approximation \cite{Y02,YBJ02}, let
us represent the atomic field operator in the form
\begin{eqnarray}
\hat{\Psi }_{a}\left( {\bf p},t\right) =&&\left\lbrack \hat{A}\left(
 {\bf p},t\right) \psi _{c}\left( p,t\right) +\hat{A}^{\dag }\left(
-{\bf p},t\right) \psi _{s}\left( p,t\right) \right\rbrack  \nonumber
\\
&&\times \exp\left( -\int\limits^{t}_{t{ } _{0}}dt_{1}\gamma |\varphi
 _{m}\left( t_{1}\right) |^{2}\right)  . \label{PsiaA}
\end{eqnarray}
The operators $\hat{A}\left( {\bf p},t\right) $ are expressible in
 terms of $\hat{\Psi }_{a}\left( {\bf p},t_{0}\right) $,
$\hat{F}\left( {\bf p},t\right) $, and the $c$-number solutions $\psi
 _{c,s}\left( p,t\right) $ of the equations
\begin{equation}
i\dot{\psi }_{c,s}\left( p,t\right) =H \psi _{c,s}\left( p,t\right)
+2g^{*}\varphi _{m}\left( t\right) \psi ^{*}_{s,c}\left( p,t\right)
 , \label{Psics}
\end{equation}
given the initial conditions $\psi _{c}\left( p,t_{0}\right) =1$,
 $\psi _{s}\left( p,t_{0}\right) =0$.

The atomic density
\begin{eqnarray}
n_{a}\left( t\right) =\left( 2\pi \right) ^{-3}\int
 d^{3}p_{1}d^{3}p_{2}\exp\left\lbrack i\left( {\bf p}_{2}-{\bf
 p}_{1}\right) \cdot {\bf r}\right\rbrack  \nonumber
\\
\times \langle \text{in}|\hat{\Psi }^{\dag }_{a}\left( {\bf
 p}_{1},t\right) \hat{\Psi }_{a}\left( {\bf p}_{2},t\right)
|\text{in}\rangle  ,
\end{eqnarray}
then appears to be ${\bf r}$-independent, and comprises the sum
$n_{a}\left( t\right) =n_{0}\left( t\right) +n_{s}\left( t\right) $
 of the densities of condensate atoms
$n_{0}\left( t\right) =|\langle $in$|\hat{\Psi }_{a}\left( 0,t\right)
 |$in$\rangle |^{2}$,  and of non-condensate (entangled) atoms
$n_{s}\left( t\right) $ in a wide spectrum of energies,
\begin{equation}
n_{s}\left( t\right) =\int dE \tilde{n}_{s}\left( E,t\right) ,
 \label{nsE}
\end{equation}
where $E$ is the kinetic energy of the non-condensate atoms.

The equation of motion for the molecular mean field $\varphi
 _{m}\left( t\right) $ is
obtained by a similar elimination of the dump fields from the
corresponding operator equation, followed by a mean-field averaging.
We thus obtain
\begin{equation}
i \dot{\varphi }_{m}\left( t\right) =g m_{a}\left( t\right) -i\left(
 \gamma n_{a}\left( t\right) +\gamma _{m}|\varphi _{m}\left( t\right)
 |^{2}\right) \varphi _{m}\left( t\right) , \label{Phim}
\end{equation}
where the parameter $\gamma _{m}$  describes molecule-molecule
 deactivation
collisions (see Ref.\ \cite{YBJW99}). Here $m_{a}$  is an anomalous
 density
containing contributions of condensate and non-condensate atoms. The
densities $m_{a}$, $n_{0}$, and $\tilde{n}_{s}$  are all expressible
 in terms of $\psi _{c}\left( p,t\right) $ and
$\psi _{s}\left( p,t\right) $. A numerical solution of Eqs.\
(\ref{Psics}) on a grid of
values of $p$, combined with Eq.\ (\ref{Phim}), is consistently
sufficient for elucidating the dynamics of the system.

The present approach becomes mathematically equivalent to the
approach of Ref.\ \cite{HPW01} if the inelastic collisions are
neglected and the detuning is time-independent. At a low molecular
density, the effect of non-condensate atoms is equivalent to the
contribution to the width of the molecular state made in Ref.\
\cite{YBJW99} for the same process.

\paragraph*{Formation of a molecular condensate.---}

Calculations were performed for two Feshbach resonances in
collisions of Na atoms, using parameter values presented in Ref.\
\cite{YBJW99}. The strong resonance, at 907 G, has the strength
 $\Delta =0.98$
G, and the weak one, at 853 G, has the strength $\Delta =9.5$ mG. The
difference of magnetic momenta $\mu =3.65$ (in Bohr magnetons), the
 elastic
scattering length $a_{a}=3.4$ nm, and the deactivation parameters are
$\gamma =0.8\times 10^{-10}$cm$^{3}/$s and $\gamma _{m}=10^{
-9}$cm$^{3}/$s. The neglection of elastic
collisions is valid whenever $n_{0}\left( t_{0}\right) \ll 10^{15}$
 cm$^{-3}$  for the weak
resonance, and $n_{0}\left( t_{0}\right) \ll 10^{17}$  cm$^{-3}$  for
 the strong one. The spatial
inhomogeneity can be neglected if the size of the condensate
substantially exceeds $8\times 10^{-2}$cm$^{-1/2}\times n^{-1
/2}_{0}\left( t_{0}\right) $ and
$2.5\times 10^{-2}$cm$^{-1/2}\times n^{-1/2}_{0}\left( t_{0}\right)
 $, respectively, for the two resonances. Even
when $n_{0}\left( t_{0}\right) =10^{8}$  cm$^{-3}$, these estimates
 set a minimal size of $8 \mu $m for
the weak resonance and $2.5 \mu $m for the strong one. The variation
 of the
magnetic field is linear in time, $B\left( t\right) =B_{0}+\dot{B} t$.

\begin{figure}

\includegraphics[width=3.375in]{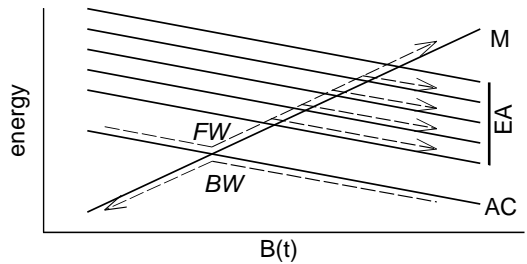}

\caption{Schematic illustration of transitions between atomic
(AC) and molecular (M) condensates and non-condensed atoms (EA) on
forward ({\it FW)} and backward ({\it BW)} sweeps.} \label{scheme}

\end{figure}
\begin{figure}

\includegraphics[width=3.375in]{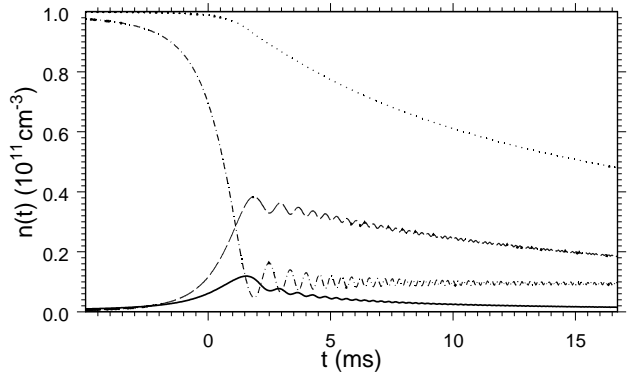}

\caption{Time dependence of the densities of the atomic
condensate (dot-dashed line), the molecular condensate (dashed line),
and the entangled atoms (solid line), calculated for the weak 853 G
resonance in Na, with the initial atomic density $n_{0}=10^{11}$
 cm$^{-3}$, and
the ramp speed $\dot{B}=-0.1$ G/s (backward sweep). The dotted line
 shows the
total atomic density (sum of the atomic densities and twice of the
molecular one).} \label{mol_td}

\end{figure}

A relatively long-lived molecular condensate is formed more
effectively in the case of a backward sweep, when the molecular state
crosses the atomic one downwards (see Fig.\ \ref{scheme}), as proposed
in Ref.\ \cite{MTJ00}. The maximal conversion efficiency of the atomic
condensate to a molecular one is $2\max\left( n_{m}\right)
/n_{0}\approx 0.8$ for the weak
resonance (see Fig.\ \ref{mol_td}). On increase of the atomic density,
or on decrease of the ramp speed, the conversion efficiency falls due
to inelastic collisions. On increase of the ramp speed, the
probability of crossing to the molecular condensate decreases, leaving
more atoms in the atomic condensate (see Ref.\ \cite{YBJW99}). At low
atomic densities the conversion becomes less efficient due to a
temporary gain of population in the non-condensate atomic states. This
observation is peculiar to, and emphasizes the importance of, the
simultaneous consideration of inelastic collisions and molecular
dissociation within the second-quantized approach. (The approach used
in \cite{YBJW99} does not describe the dissociation in the backward
sweep.)

\begin{figure}

\includegraphics[width=3.375in]{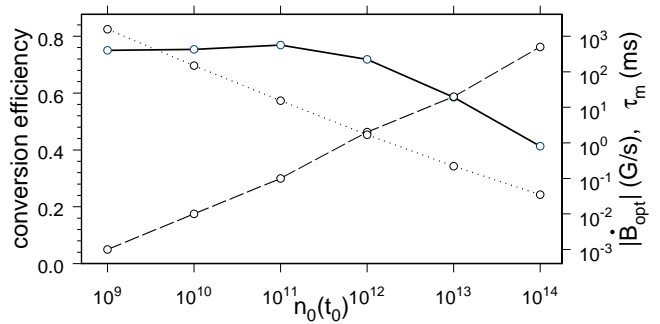}

\caption{Conversion efficiency (solid line), optimal ramp speed
$\dot{B}_{\text{opt}}$  (dashed line), and lifetime of the molecular
 condensate $\tau _{m}$
(dotted line) as a function of the initial atomic density, calculated
for the weak resonance in Na (in a backward sweep).} \label{mol_den}

\end{figure}

Figure \ref{mol_den} shows that a substantial conversion
efficiency is retained in a wide region of the condensate density,
leaving much freedom in the choice of the ramp speed appropriate for
experiments. The lesser the initial density, the longer the lifetime
$\tau _{m}$  of the molecular condensate but the higher the precision
 required
for the control of the magnetic field.

The optimal ramp speed is approximately proportional to the
initial density. This dependence minimizes the effect of a variation
of parameters determining the conversion of the atomic condensate to
the molecular one and loss of the molecular condensate. Indeed, the
conversion to the molecular condensate is (in the fast decay
approximation \cite{YBJW99}) characterized by the parameter
$g^{2}n_{0}/\dot{B}$. Similarly, the loss is characterized by the
 ratio of the
deactivation lifetime (which is inversely proportional to the initial
density), and the crossing time (which is inversely proportional to
the ramp speed).

The use of the strong resonance achieves a lower conversion
efficiency, due to a gain in the temporary formation of non-condensate
atoms. The optimal ramp speed is more than two orders of magnitude
larger than in the weak resonance, given the same initial density.

\paragraph*{Formation of an entangled gas.---}

As demonstrated in Ref.\ \cite{YBJ02}, the non-condensate atoms
are formed in squeezed states, which now turn out to be two-mode
squeezed states, as in Ref.\ \cite{Y02}. It is similar to the state of
electromagnetic radiation formed by parametric down conversion. As in
quantum optics Ref.\ \cite{SZ97}, the amount of squeezing can be
measured by the energy-dependent parameter $r\left( E,t\right) $
 related to the
maximal and minimal uncertainties of quadratures involving creation
and annihilation operators for opposite momenta \cite{Y02}. A mean
squeezing parameter, weighed by the spectral density of Eq.\
(\ref{nsE}),
\begin{equation}
\bar{r}\left( t\right) =\int dE \tilde{n}_{s}\left( E,t\right) r\left
( E,t\right) /n_{s}\left( t\right) . \label{avsq}
\end{equation}
is used to describe the time variation of the squeezing.

\begin{figure}

\includegraphics[width=3.375in]{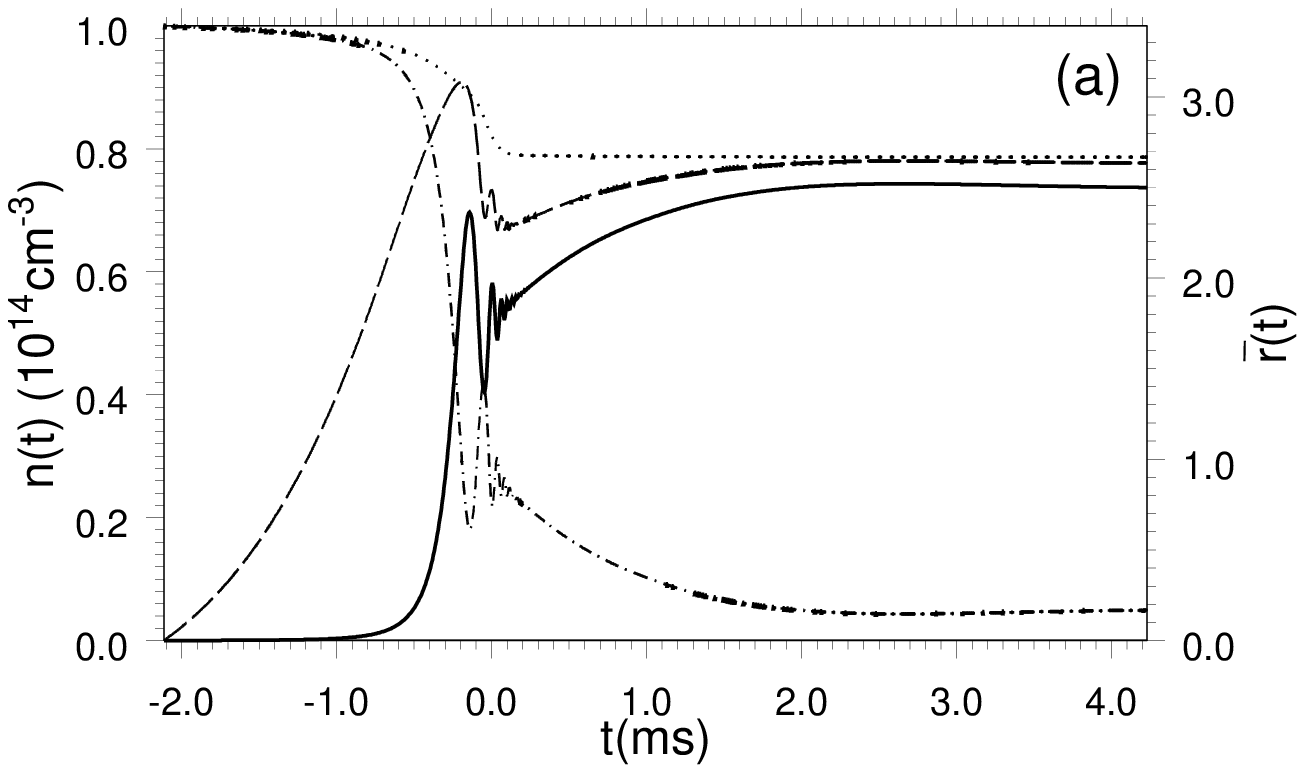}
\includegraphics[width=3.375in]{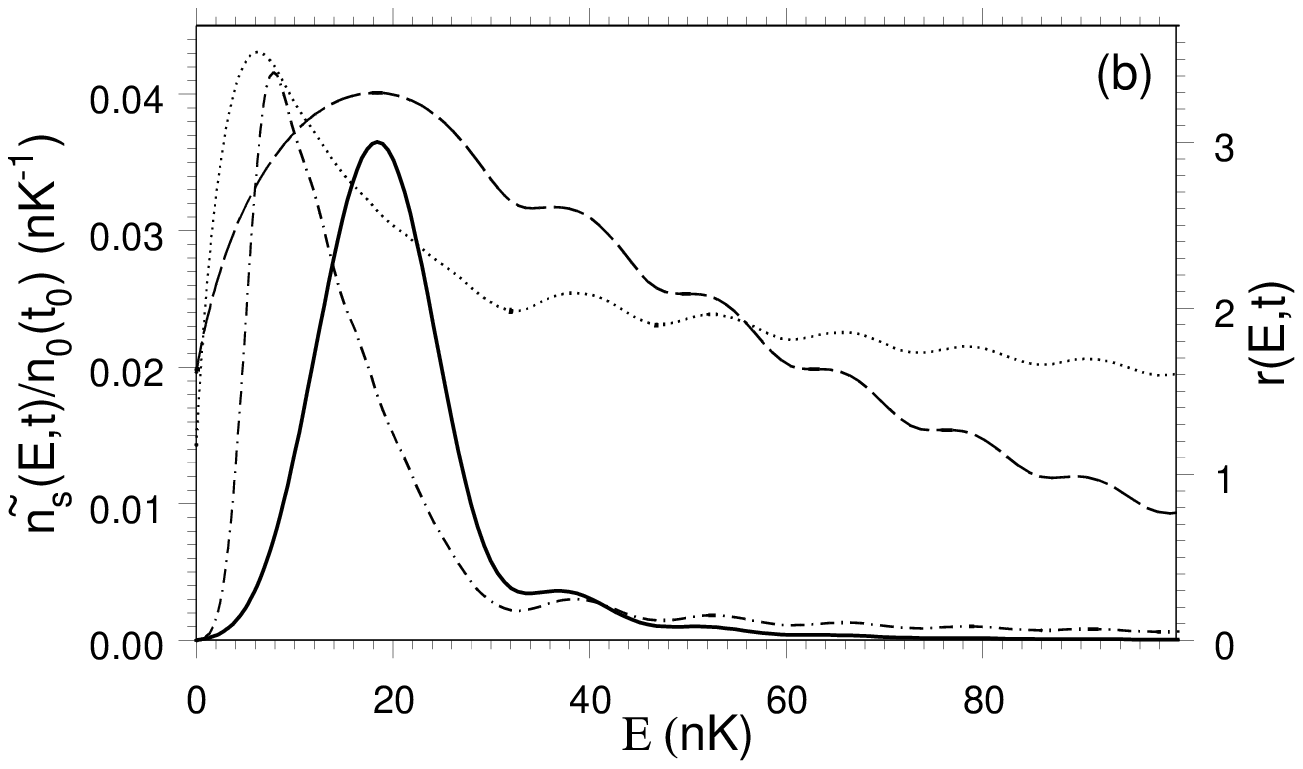}

\caption{(a) Time dependence of the densities of the atomic
condensate (dot-dashed line), entangled atoms (solid line), and the
total atomic density (dotted line) calculated for the weak resonance
in Na with the initial atomic density $n_{0}=10^{14}$  cm$^{-3}$  and
 ramp speed 50
G/s in a forward sweep. The dashed line shows the mean squeezing
parameter $\bar{r}\left( t\right) $ [see Eq.\ (\protect\ref{avsq})].
(b) Energy spectra of
the entangled-atom density $\tilde{n}_{s}\left( E,t\right) $ (solid
 line) and the squeezing
parameter $r\left( E,t\right) $ (dashed line) calculated at the peak,
 $t\approx -0.19$ ms. The
dot-dashed and dotted lines show their values on the plateau at
 $t\approx 4$
ms.} \label{ent_sp}

\end{figure}

A stable gas of entangled atoms is formed by a forward sweep, in
which the molecular state crosses the atomic one upwards (see Fig.\
\ref{scheme}). This process, too, is more efficient in the weak
resonance. The molecular density is then very low and persists a
shorter time (compared to that in the backward sweep) due to fast
dissociation. Figure \ref{ent_sp}a demonstrates that more than 70\% of
the atomic condensate can be transformed into a gas of atoms in two-
mode squeezed states with the mean squeezing parameter
 $\bar{r}\approx 2.6$,
corresponding to a noise reduction of about 23 dB. The time dependence
of the mean squeezing has a peak of $\bar{r}\approx 3.1$ at $t\approx
 -0.19$ ms. The state of
an entangled gas can be freezed at the peak time by fast turning off
of the magnetic field. The energy spectra of the entangled-atom
density and the squeezing parameter are presented in Fig.\
\ref{ent_sp}b. The density spectra are rather narrow, and the peak
energy increases with time. The squeezing parameter reaches the value
of $r\left( E,t\right) \approx 3.5$ (noise reduction of 30 dB) at the
 energy $E\approx 6$ nK and the
time $t\approx 4$ ms.

\paragraph*{Conclusions.---}

Both quantum corrections and deactivating collisions are
necessary for the analysis of association in atomic BEC due to
Feshbach resonance in a time-dependent magnetic field. Over 80\% of
 the
atomic population can be converted to a molecular condensate in a
backward sweep. The molecules dissociate onto atoms in two-mode
squeezed states that are entangled. In a forward sweep, very high
squeezing may be obtained with the parameter $r$ reaching a value of
more than 3.

\end{document}